\documentclass[useAMS,usenatbib]{mn2e}
\usepackage{graphicx}
\usepackage{txfonts}

        \title[The fraction of Compton-thick sources in an INTEGRAL complete AGN sample]
  {The fraction of Compton-thick sources in an INTEGRAL complete AGN sample}
\author[A. Malizia et al.]
{A.~Malizia$^1$,
  J.~B.~Stephen$^1$, L.~Bassani,$^1$, A.~J.~Bird$^2$, F.~Panessa$^3$, P.~Ubertini$^3$\\
$^1$ IASF/INAF, via Gobetti 101, I-40129 Bologna, Italy \\
$^2$ School of Physics and Astronomy, University of Southampton, SO17 1BJ, Southampton, U.K.\\
$^3$ IASF/INAF, via del Fosso del Cavaliere 100, I-00133 Roma, Italy }

\begin{document}

\date{}

\pagerange{\pageref{firstpage}--\pageref{lastpage}} \pubyear{2007}
\maketitle
\label{firstpage}
\begin{abstract}
 We study the N$_{H}$ distribution in a complete sample of 88 AGN selected in the 20-40 keV band from INTEGRAL/IBIS observations.
 We find that the fraction of absorbed (N$_{H}$ $\geq$ 10$^{22}$ cm$^{2}$) sources is 43\% while the Compton thick AGN comprise 
7\% of the sample. While these estimates are fully compatible with previous soft gamma-ray surveys, 
they would appear to be in contrast with results reported by Risaliti et al. (1999) using an optically selected sample.
This apparent difference can be explained as being due to a selection bias caused by the reduction in high energy flux in 
Compton thick objects rendering them invisible at our sensitivity limit. Taking this into account we estimate that the 
fraction of highly absorbed sources is actually in close agreement with the optically selected sample. 
Furthermore we show that the measured fraction of absorbed sources in our sample decreases from 80\% to 
$\sim$ 20-30\% as a function of redshift with all Compton thick AGN having $z$ $\leq$ 0.015. 
If we limit our analysis to this distance and compare only the type 2 objects in our sample with the Risaliti et al.  
objects below this redshift value, we find a perfect match to their N$_{H}$ distribution.
We conclude that in the low redshift bin we are seeing almost the entire AGN population, from unabsorbed to at least mildly 
Compton thick objects, while in the total sample we lose the heavily absorbed 'counterparts' of distant and therefore 
dim sources with little or no absorption. 
Taking therefore this low z bin as the only one able to provide the 'true' distribution of absorption in type 1 and 2 AGN, 
we estimate the  fraction of Compton thick objects to be $\geq$24\%.

  \end{abstract}


\section{Introduction}
Both direct and indirect evidence suggests that a large fraction of AGN are obscured in X-rays by large
amounts of gas and dust which prevents their nuclei being seen up to the energy, which depends on the column density of the absorber, 
at which the obscuring material becomes transparent. 
So far, X-ray observations below 10 keV have extensively probed the so called Compton thin regime, i.e. column
densities below 1.5 $\times$ 10$^{24}$  cm$^{-2}$ (the inverse of the Thomson cross-section) but still in excess of the
Galactic value in the source direction. The Compton thick regime has been much less sampled either due
to the lack of complete spectral coverage and/or all-sky surveys above 10 keV (for mildly Compton thick
sources) or because the entire high energy spectrum is down scattered by Compton recoil and therefore
depressed at all energies (heavily Compton thick sources). Until now, indirect arguments have been used to
probe this regime: the intensity of the iron line at 6.4 keV (equivalent width typically of the order of 1 keV, Matt 1999),
the signature of strong Compton reflection, or the ratio of the observed X-ray luminosity against an isotropic
indicator of the source intensity, often the [OIII]5007$\AA$ luminosity. However, sometimes iron line and Compton
reflection diagnostics may lead to a wrong classification, caused by a temporary switching off of the primary
continuum (Guainazzi et al. 2005) and not by thick absorption. Furthermore, the [OIII] luminosity is not
always available and/or properly estimated so that the large uncertainties on the L$_{X}$/L$_{[OIII]}$ ratios can also
lead to a misclassification. \\
The study of Compton thick AGN is important for various reasons:
 (i) about  80\% of the active galactic nuclei in the local Universe are obscured (e.g., Maiolino et al. 1998;
Risaliti et al. 1999); 
(ii) their existence is postulated in all AGN synthesis models of the X-ray background (Gilli et al 2007);
(iii) they may constitute an important ingredient for the IR and the sub-mm backgrounds, where most of
the absorbed radiation is re-emitted by dust (Fabian \& Iwasawa 1999; Brusa et al. 2001) and
(iv) accretion in these objects may contribute to the local black hole mass density (Fabian \& Iwasawa 1999, Marconi et al. 2004).\\
Because of this interest and despite the limitations so far encountered, a sizable sample of Compton
thick AGN is available for in depth studies (Della Ceca  et al. 2008). However, this sample is by no means complete,
properly selected and reliable in relation to the column density estimates. It is clear that for an unbiased
census of Compton thick sources sensitive soft gamma-ray surveys/observations are needed.\\
A step forward in this field is now provided by Swift/BAT and INTEGRAL/IBIS which are surveying the sky above 20 keV with a
sensitivity better than a few mCrab and a point source location accuracy of 1-3 arcmin depending on the
source strength and distance (Bird et al. 2007). These two surveys are complementary, not only because they
probe the sky in a different way but also because they can be a check of each other's results. Together they
will provide the best yet knowledge of the extragalactic sky at gamma-ray energies.
Results obtained so far from these two instruments, point to a percentage  of absorbed sources (N$_{H}>$ 10$^{22}$  
cm$^{-2}$) in the range 50-65\%, while the fraction of Compton thick objects is constrained  to be $<$ 20\%, likely closer to 10\% (Ajello 2009).
Synthesis models of the cosmic X-ray background  predict instead a fraction of Compton thick AGN close to 50\% (Gilli et al. 2007) although
recent revisions seem to suggest a smaller contribution of 9\% (Triester et al. 2009).\\
In this work we use a complete sample of INTEGRAL selected AGN to study the distribution of  the absorption in the
local Universe. While we find that the overall picture is in agreement with previous hard X-ray survey results, a more in depth 
analysis of our sample  shows that  80\% of the sources are absorbed  and 24\%, or even more, are Compton thick.

\begin{table*}
\begin{center}
\centerline{Table 1: INTEGRAL/IBIS complete sample of AGN}
\label{table:1}
 \begin{tabular}{r l l l  c l l l c r l}
  \hline\hline
  &  Name              &      RA     & Dec           &  Class   &  $z$   & F$_{20-100 keV}^{\dagger}$  & F$_{2-10 keV}^{\dagger}$ & Log N$_H$ & Ref \\
\hline
1  & IGR J00333+6122    & 00 33 18.41 & +61 27 43.1  &  Sy 1.5  & 0.1050 &                  1.189      &         0.679            & 21.93 & 1  \\
2  & 1ES 0033+595       & 00 35 52.64 & +59 50 04.6  &  BL Lac  & 0.0860 &                  1.869      &         5.900            & 21.56 & 2 \\
3  & NGC 788            & 02 01 06.40 & -06 48 56.0  &  Sy 2    & 0.0136 &                  5.021      &         0.468            & 23.48 & 3 \\
4  & NGC 1068           & 02 42 40.71 & -00 00 47.8  &  Sy 2    & 0.0038 &                  3.001      &         0.5$^{\star}$    & 25.00 & 4  \\
5  & QSO B0241+62       & 02 44 57.69 & +62 28 06.5  &  Sy 1    & 0.0440 &                  6.342      &         1.840            & 21.50 & 5\\
6  & NGC 1142           & 02 55 12.19 & -00 11 02.3  &  Sy 2    & 0.0288 &                  5.247      &         0.960            & 23.65 & 6 \\
7  & B3 0309+411B       & 03 13 01.96 & +41 20 01.2  &  Sy 1    & 0.1360 &                  2.569      &         2.830            & {\bf 21.10} & 5 \\
8  & IGR J03184-0014    & 03 18 28.91 & -00 15 23.1  &  QSO     & 1.9820 &                  4.949      &          -               &       -     & - \\
9  & NGC 1275           & 03 19 48.16 & +41 30 42.1  &  Sy 2    & 0.0175 &                  3.759      &         1.23             & 21.08 & 7 \\
10  & 3C 111             & 04 18 21.28 & +38 01 35.8  &  Sy 1    & 0.0485 &                  9.455      &         6.800            & 21.63 & 5 \\
11 & LEDA 168563        & 04 52 04.85 & +49 32 43.7  &  Sy 1    & 0.0290 &                  6.305      &         4.540            & {\bf 21.73} & 6 \\
12 & 4U 0517+17         & 05 10 45.51 & +16 29 55.8  &  Sy 1.5  & 0.0179 &                  6.305      &         2.530            & 20.95 & 1 \\
13 & MCG+08-11-11       & 05 54 53.61 & +46 26 21.6  &  Sy 1.5  & 0.0205 &                  6.251      &         5.620            & {\bf 21.32} &  1\\
14 & Mkn 3              & 06 15 36.36 & +71 02 15.1  &  Sy 2    & 0.0135 &                 10.115      &         0.650            & 24.00 & 8\\
15 & Mkn 6$^{\ddagger}$ & 06 52 12.25 & +74 25 37.5  &  Sy 1.5  & 0.0188 &                  5.019      &         2.510            & 22.68 & 9\\
16 & IGR J07565-4139    & 07 56 19.71 & -41 37 41.6  &  Sy 2    & 0.0210 &                  1.756      &         0.150            & 21.77 & 6 \\
17 & IGR J07597-3842    & 07 59 41.66 & -38 43 57.3  &  Sy 1.2  & 0.0400 &                  3.832      &         2.370            & {\bf 21.78} & 1 \\
18 & ESO 209-12         & 08 01 57.60 & -49 46 42.0  &  Sy 1    & 0.0396 &                  2.283      &         0.830            & {\bf 21.38} & 10 \\
19 & FRL 1146           & 08 38 30.78 & -35 59 33.4  &  Sy 1.5  & 0.0316 &                  1.756      &         1.18$^{\star}$   & 21.50 & 10\\
20 & QSO B0836+710      & 08 41 24.36 & +70 53 42.2  &  Blazar  & 2.1720 &                  5.546      &         2.630            & {\bf 20.47} & 2\\
21 & SWIFT J0917.2-6221 & 09 16 09.01 & -62 19 29.0  &  Sy 1    & 0.0573 &                  2.172      &         2.170            & 21.67 & 6 \\
22 & MCG-05-23-16       & 09 47 40.15 & -30 56 55.9  &  Sy 2    & 0.0085 &                 14.500      &         8.200            & 22.18 & 11\\
23 & IGR J09523-6231    & 09 52 20.50 & -62 32 37.0  &  Sy 1.9  & 0.2520 &                  1.246      &         0.910            & 22.90 & 12\\
24 & SWIFT J1009.3-4250 & 10 09 48.12 & -42 48 42.6  &  Sy 2    & 0.0330 &                  2.870      &         0.200            & 23.43 & 13\\
25 & NGC 3281           & 10 31 52.06 & -34 51 13.3  &  Sy 2    & 0.0115 &                  5.132      &         0.290            & 24.3  & 14\\
26 & SWIFT J1038.8-4942 & 10 38 44.87 & -49 46 52.7  &  Sy 1.5  & 0.0600 &                  1.453      &         1.450            & 21.79 & 6 \\
27 & IGR J10404-4625    & 10 40 22.27 & -46 25 24.7  &  Sy 2    & 0.2392 &                  3.228      &         0.720            & 22.43 & 6\\
28 & NGC 3783           & 11 39 01.72 & -37 44 18.9  &  Sy 1    & 0.0097 &                 13.412      &         6.030            & 22.06 & 1\\
29 & IGR J12026-5349    & 12 02 47.63 & -53 50 07.7  &  Sy 2    & 0.0280 &                  3.965      &         0.800            & 22.52 & 3\\
30 & NGC 4151$^{\ddagger}$ & 12 10 32.58 & +39 24 20.6  &  Sy 1.5  & 0.0033 &                 63.379      &        25.000         & 23.34 & 1\\
31 & 4C 04.42           & 12 22 22.55 & +04 13 15.8  &  QSO     & 0.9650 &                  2.641      &         0.250            & {\bf 20.23} & 15 \\
32 & Mkn 50             & 12 23 24.14 & +02 40 44.8  &  Sy 1    & 0.0234 &                  1.398      &         0.980            & $<$ 21.08 & 1\\
33 & NGC 4388           & 12 25 46.75 & +12 39 43.5  &  Sy 2    & 0.0084 &                 23.705      &         1.700            & 23.52 & 16\\
34 & 3C 273             & 12 29 06.70 & +02 03 08.6  &  QSO     & 0.1583 &                 18.479      &         9.62$^{\star}$   & {\bf 20.23} & 2 \\
35 & NGC 4507           & 12 35 36.62 & -39 54 33.4  &  Sy 2    & 0.0118 &                 17.382      &         2.100            & 23.46 & 17\\
36 & LEDA 170194        & 12 39 06.32 & -16 10 47.8  &  Sy 2    & 0.0360 &                  6.15       &         1.11             & 22.49 & 6 \\
37 & NGC 4593           & 12 39 39.42 & -05 20 39.3  &  Sy 1    & 0.0090 &                  7.230      &         3.720            & {\bf 20.30} & 1 \\
38 & IGR J12415-5750    & 12 41 25.36 & -57 50 03.9  &  Sy 1    & 0.0230 &                  2.284       &         0.770            & {\bf 21.48} & 1 \\
39 & 3C 279             & 12 56 11.17 & -05 47 21.5  & Blazar   & 0.5362 &                  2.277      &         0.60$^{\star}$   & {\bf 20.30} & 2\\
40 & NGC 4945           & 13 05 27.48 & -49 28 05.6  &  Sy 2    & 0.0019 &                 29.873      &         0.355            & 24.60 & 17\\
41 & IGR J13091+1137    & 13 09 05.60 & +11 38 02.9  &  Sy 2    & 0.0251 &                  4.549      &         0.251            & 23.95 & 18\\
42 & IGR J13109-5552    & 13 10 43.08 & -55 52 11.7  &  Sy 1    & 0.0850 &                  2.416      &         0.510            & $<$ 21.66 & 1 \\
43 & Cen A              & 13 25 27.61 & -43 01 08.8  &  Sy 2    & 0.0018 &                 74.434      &         8.510            & 23.54  & 17\\
44 & MCG-06-30-15       & 13 35 53.78 & -34 17 44.1  &  Sy 1.2  & 0.0077 &                  4.685      &         3.460            & 22.17  & 1\\
45 & NGC 5252           & 13 38 16.00 & +04 32 32.5  &  Sy 2    & 0.0230 &                  4.628      &         3.000            & 22.83  & 19 \\  
46 & 4U 1344-60$^{\ddagger}$ & 13 47 36.00 & -60 37 04.0  &  Sy 1.5  & 0.0130 &                  7.399      &         3.540       & 23.63  & 10\\
47 & IC 4329A           & 13 49 19.26 & -30 18 34.0  &  Sy 1.2  & 0.0160 &                 21.518      &        10.400            & 21.54  & 1\\
48 & Circinus Galaxy    & 14 13 09.91 & -65 20 20.5  &  Sy 2    & 0.0014 &                 21.695      &         1.000            & 24.60  & 17\\
49 & NGC 5506           & 14 13 14.87 & -03 12 26.9  &  Sy 1.9  & 0.0062 &                  8.820      &         8.380            & 22.53  & 17\\
50 & ESO 511-G030       & 14 19 22.66 & -26 38 34.4  &  Sy 1    & 0.2239 &                  3.455      &         1.30             & {\bf 20.7} & 20 \\
51 & IGR J14515-5542    & 14 51 33.43 & -55 40 39.4  &  Sy 2    & 0.0180 &                  1.623      &         0.710            & 21.59  & 6\\
52 & IC 4518A           & 14 57 41.18 & -43 07 55.6  &  Sy 2    & 0.0163 &                  2.247      &         0.617            & 23.15  & 3\\
53 & IGR J16024-6107    & 16 01 48.40 & -61 08 53.6  &  Sy 2    & 0.0110 &                  1.152      &         0.180            & {\bf 21.45}  &  21 \\
54 & IGR J16119-6036    & 16 11 51.36 & -60 37 53.1  &  Sy 1    & 0.0160 &                  2.548      &         0.330            & {\bf 21.36}  & 1\\
55 & IGR J16185-5928    & 16 18 25.68 & -59 26 45.6  &  NLS1    & 0.0350 &                  1.850      &         0.50$^{\star}$   & {\bf 21.39}  & 22\\
56 & IGR J16351-5806    & 16 35 13.42 & -58 04 49.7  &  Sy 2    & 0.0091 &                  1.982      &         0.031            &  24.57  & 23 \\
57 & IGR J16385-2057    & 16 38 30.91 & -20 55 24.6  &  NLS1    & 0.0269 &                  1.625      &         0.700            & {\bf 21.08}  & 22 \\
58 & IGR J16426+6536    & 16 43 04.07 & +65 32 50.9  &  NLS1    & 0.3230 &                  4.325      &          -               &   -          &  - \\
\hline
\end{tabular}
\end{center}
\end{table*}

\begin{table*}
\begin{center}
\centerline{Table 1: continued }
\label{table:2}
 \begin{tabular}{l l l  c l l l c r l }
  \hline\hline
 & Name              &      RA     & Dec            &  Class   &  $z$   & F$_{20-100 keV}^{\dagger}$  & F$_{2-10 keV}^{\dagger}$ & Log N$_H$  & Ref \\
\hline

59 & IGR J16482-3036     & 16 48 14.94 & -30 35 06.1  &  Sy 1    & 0.0310 &                  3.077      &         2.000            & 21.00 & 1\\
60 & IGR J16558-5203$^{\ddagger}$     & 16 56 05.73 & -52 03 41.2  &  Sy 1.2  & 0.0540 &                  3.341      &         1.770            & 23.27 & 10 \\ 
61 & SWIFT J1656.3-3302  & 16 56 16.83 & -33 02 12.2  &  Blazar  & 2.4000 &                  2.321      &         0.440            & {\bf 21.34} & 24\\
62 & NGC 6300            & 17 16 59.47 & -62 49 14.0  &  Sy 2    & 0.0037 &                  6.722      &         1.290            & 23.36  & 19\\
63 & GRS 1734-292        & 17 37 28.35 & -29 08 02.5   &  Sy 1    & 0.0214 &                  8.402      &         3.840            & $>$21.32 & 1\\
64 & 2E 1739.1-1210      & 17 41 55.30 & -12 11 57.0   &  Sy 1    & 0.0370 &                  2.755      &         1.290            & 21.18 & 1 \\
65 & IGR J17488-3253     & 17 48 54.82 & -32 54 47.8   &  Sy 1    & 0.0200 &                  3.963      &         1.400            & 21.53 & 1 \\
66 & IGR J17513-2011     & 17 51 13.62 & -20 12 14.6   &  Sy 1.9  & 0.0470 &                  2.737      &         0.548            & 21.85 & 25 \\
67 & IGR J18027-1455     & 18 02 45.50 & -14 54 32.0   &  Sy 1    & 0.0350 &                  4.342      &         0.660            & 21.48 & 1 \\
68 & IGR J18249-3243     & 18 24 56.11 & -32 42 58.9   &  Sy 1    & 0.3550 &                  1.121      &         0.520            & {\bf 21.14} & 26\\
69 & IGR J18259-0706     & 18 25 57.58 & +07 10 22.8   &  Sy 1    & 0.0370 &                  1.605      &         0.540            & 22.03 & 1 \\
70 & PKS 1830-211        & 18 33 39.89 & -21 03 39.8   &  Blazar  & 2.5070 &                  5.171      &         1.000            & 22.00 & 27\\
71 & ESO 103-35          & 18 38 20.30 & -65 25 41.0   &  Sy 2    & 0.0133 &                  8.439      &         2.500            & 23.30  & 28 \\
72 & 3C 390.3            & 18 42 08.99 & +79 46 17.1   &  Sy 1    & 0.0561 &                  6.057      &         3.350            & {\bf 20.63} & 5\\
73 & 2E 1853.7+1534      & 18 56 01.28 & +15 38 05.7   &  Sy 1    & 0.0840 &                  2.454      &         1.220            & $<$21.59 & 1\\
74 & IGR J19378-0617     & 19 37 39.00 & -06 13 06.0   &  NLS1    & 0.0106 &                  1.889      &         3.500            & {\bf 21.17} & 22 \\
75 & NGC 6814            & 19 42 40.64 & -10 19 24.6   &  Sy 1.5  & 0.0052 &                  5.908      &         0.17$^{\star}$   & {\bf 21.10} & 1\\
76 & Cyg A               & 19 59 28.36 & +40 44 02.1   &  Sy 2    & 0.0561 &                  8.155      &         1.820            & 23.57 & 17\\
77 & IGR J20186+4043     & 20 18 38.73 & +40 40 59.9   &  Sy 2    & 0.0144 &                  2.209      &         0.430            & 23 & 29 \\
78 & 4C 74.26            & 20 42 37.30 & +75 08 02.4   &  Sy 1     & 0.1040 &                  4.872      &         2.530            & 21.15 & 1\\
79 & S52116+81           & 21 14 01.18 & +82 04 48.3   &  Sy 1    & 0.0840 &                  4.059      &         1.210            & $<$21.38 & 1\\
80 & IGR J21247+5058$^{\ddagger}$ & 21 24 39.44 & +50 58 24.4   &  Sy 1    & 0.0200 &                 10.853      &         4.880   & 22.89 & 1 \\
81 & SWIFT J2127.4+5654  & 21 27 45.58 & +56 56 35.6   &  NLS1    & 0.0140 &                  2.683      &         1.890            & {\bf 21.90} & 22\\
82 & RX J2135.9+4728     & 21 35 54.02 & +47 28 22.3   &  Sey 1   & 0.0250 &                  1.605      &         0.770            & 21.36 & 30 \\                 
83 & NGC 7172            & 22 02 01.91 & -31 52 11.3   &  Sy 2    & 0.0087 &                  8.249      &         1.300            & 22.93 & 12 \\
84 & BL LAC              & 22 02 43.29 & +42 16 40.0   &  BL Lac  & 0.0686 &                  2.661      &         2.020            & {\bf 21.44} & 2 \\
85 & MR 2251-178         & 22 54 05.80 & -17 34 55.0   &  Sy 1    & 0.0640 &                  7.284      &         2.000            & 22.33 & 1\\
86 & MCG-02-58-22        & 23 04 43.47 & -08 41 08.6   &  Sy 1.5  & 0.0469 &                  3.909      &         3.180            & {\bf 20.56} & 1\\
87 & IGR J23308+7120     & 23 30 37.28 & +71 22 46.6   & Sy 2     & 0.037  &                  1.246      &         0.140            & 22.78 & 21\\
88 & IGR J23524+5842     & 23 52 22.11 & +58 45 30.7   &  Sy 2    & 0.1640 &                  1.280      &         0.290            & 22.80 & 12\\
\hline
\hline
\end{tabular}
\end{center}
\small
Notes: in {\bf bold} upper limit values (Galactic column densities) of N$_{H}$; $\dagger$: $\times$ 10$^{-11}$  s$^{-1}$ cm$^{-2}$; 
$^{\ddagger}$ source with complex absorption modelled with \texttt{pcfabs} (partial covering) in XSPEC for which the highest value of column density 
has been reported; $^{\star}$ 2-10 keV  flux variation. Ref: (1) Molina et al. 2009; (2) Donato et. al. 2005; (3) De Rosa et al. 2008a; 
(4) Matt et al. 2005, (5) Molina et al. 2008; (6) Malizia et al. 2007; (7) Churazov et al.  2003;
(8) Cappi et al. 1999; (9) Malizia et al. 2003; (10) Panessa et a. 2008; (11) Braito et al. 2007;
(12) Rodriguez et al. 2008; (13) Landi et al. 2007a; (14) Vignali \& Comastri 2002; (15) De Rosa et al. 2008b; 
(16) Panessa 2003; (17) Bassani et al. 1999; (18) Sazonov et al. 2005; 
(19) Risaliti 2002; (20) ASCA Tartarus Archive; (21) Landi et al. 2007b; (22) Malizia et al. 2008; 
(23) Malizia et al. 2009;  (24) Masetti et al. 2008a; (25) XMM-Newton Archive; (26) Landi et al. 2007c; (27) De Rosa et al. 2005; 
(28) Wilkes et al. 2001; (29) Pandel et al. 2008; (30) Swift-XRT Archive.
\end{table*}

\section{The INTEGRAL/IBIS complete sample of AGN}
The complete sample of INTEGRAL selected AGN has been extracted from a set of 140 extragalactic objects detected in the
20-40 keV band and listed in the 3$^{rd}$ IBIS survey (Bird et al. 2007).
Most of these objects were already identified as active galaxies in the IBIS catalogue, while others were subsequently 
classified as such thanks to follow-up  optical spectroscopy.\footnote{For optical
classification of \emph{INTEGRAL} sources, please refer to Masetti's web page at http://www.iasfbo.inaf.it/extras/IGR/main.html}\\
>From this list, a complete sample has been extracted by means of the V/V$_{max}$ test, which was first introduced by Shmidt (1968) as a test
of uniformity of distribution in space for a flux-limited sample of
objects. It can, however, be used in the opposite sense, that is,
assuming that the sample is distributed uniformly in space (and that
there is no evolution), it is possible to test if the sample is
complete. The test consists of comparing the volumes contained within
the distances where the sources are observed (V) with the maximum
volumes (V$_{max}$), defined as those within the distance at which
each source would be at the limit of detection. If the sample is not
complete, the expected value for $<$V/V$_{max}$$>$ is less than 0.5,
while when complete it should be equal to 0.5.

\begin{figure}
\centering
\includegraphics[width=0.8\linewidth]{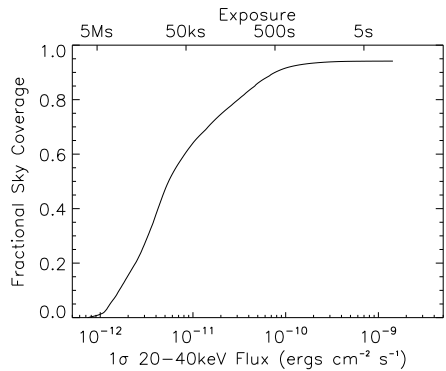}
\caption{{\small The fraction of the sky seen as a function of both 1$\sigma$ limiting flux and exposure
for the complete 3$^{rd}$ catalogue. It can be seen that large fractions of the sky have very different sensitivity limits.}}
\label{fig1}
\end{figure}

\begin{figure}
\centering
\includegraphics[width=0.8\linewidth]{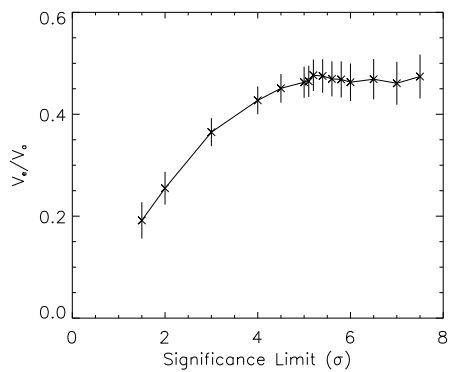}
\caption{{\small The value of $<$V$_e$/V$_a$$>$ as a function of limiting significance. }}
\label{fig2}
\end{figure}

In the case of the IBIS catalogue, the sky exposure, and therefore
the limiting sensitivity is a strong function of position, as is shown in  figure 1.
This can be taken into consideration by using the V$_e$/V$_a$ variation of the test, introduced by
Avni \& Bahcall (1980).  Once again the expected mean value
m=$<$V$_e$/V$_a$$>$ will be 0.5 when the sample is complete.\\
For our specific case, the significance for each source given
in the catalogue is not that found in the sky map, but a value which
is adjusted after the source is detected and a light curve created for
it. In applying the V$_e$/V$_a$ test, the significances used are those
which are the basis of finding the source i.e. from the sky map.
Figure 2 shows the value of $<$V$_e$/V$_a$$>$ as a function of
limiting sensitivity. It can be seen that the increasing trend becomes flat above about 5.2$\sigma$ 
at which point the ratio has a value of 0.47$\pm$0.03, consistent with completeness. \\
There are 88 objects detected in the 20-40 keV band  with a significance higher than this limit and 
they form our complete sample of INTEGRAL selected AGN:  46 objects are  of type 1 (Seyfert 1-1.5, of which 
5 Narrow Line Seyfert 1s) and  33 of type 2 (Seyfert 1.8-2); only 9 Blazars  (BL Lac-QSO) are included in the catalogue. 
Table 1 lists all our  objects together with their coordinates  and  optical class, redshift, gamma-ray  
(20-100 keV) and X-ray (2-10 keV) fluxes, N$_{H}$ value and relevant reference.
In those sources where the intrinsic absorption is  lower than or  equal to the Galactic value, 
this has been listed and  highlighted in the table  (bold values); 
in the following these values will be treated as upper limits on N$_{H}$.
While the 20-100 keV fluxes are taken from the INTEGRAL survey (Bird et al. 2007), the 2-10 keV fluxes and column densities have been 
collected from the literature (see reference in column 10 of table 1) with the exception of IGR J17513-2011 and RX J2135.9+4728. 
These two sources have  been observed by XMM-Newton and Swift-XRT respectively;
their X-ray data, never published, have been analysed in the present work and the resulting X-ray fluxes and column densities are reported 
here for the first time. 
Only two objects (the QSO, IGR J03184-0014 and the NLS1, IGR J16426+6536) have no X-ray data available and hence no 
column density estimate; given their optical classification (which suggests that they are not heavily absorbed objects) 
and that there are only two, we assume that neglecting them from the following discussion will not alter the main conclusion of this work.\\
The luminosities have been calculated for all  sources  
assuming H$_{0}$=71 km s$^{-1}$ Mpc$^{-1}$ and q$_{0}$ = 0;  
in figure 3  the Log (L$_{20-100~keV}$) is plotted against the redshifts to show  the large range of 
parameters sampled by the present work. 
From this figure it can be estimated that  our sensitivity limit is around  1.5 $\times$ 10$^{-11}$
erg cm$^{-2}$ s$^{-1}$.  We find that the redshift spans from 0.0014 to 2.5 with a mean at 0.161 while the 
luminosity ranges from Log (L$_{20-100~keV}$)  $\sim$ 42 to almost Log (L$_{20-100~keV}$) $\sim$ 48 with a mean at  
Log (L$_{20-100~keV}$) $\sim$ 44.

\begin{figure}
\centering
\includegraphics[width=0.8\linewidth]{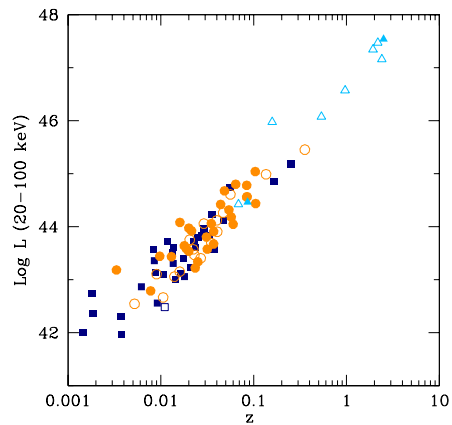}
\caption{Hard X-ray luminosity vs. redshift for all the complete AGN sample. Circles are type 1 objects, squares are type 2 and triangles 
are blazars. Open symbols are objects where no intrinsic absorption have been measured. }
\label{fig3}
\end{figure}

\begin{figure}
\centering
\includegraphics[width=0.8\linewidth]{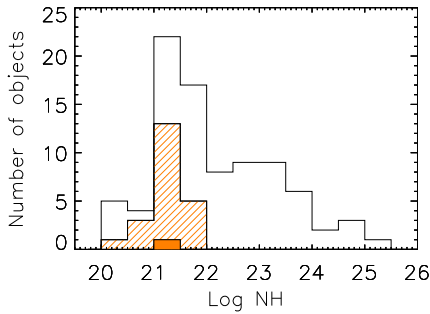}
\caption{Distribution of column density in the INTEGRAL complete sample. The dashed bins represent
upper limit measurements (including Galactic values, see text), while the filled bin corresponds to GRS 1734-292 
for which a lower limit is available.}\label{nh_tot}
\end{figure}

\section{Column Densities in our sample}
The column density distribution for the complete sample is shown in figure 4. Assuming N$_{H}$ = 10$^{22}$ cm$^{-2}$
as the dividing line between absorbed and unabsorbed sources, we find that absorption is present in 43$\%$ of the sample.  
Within our catalogue we find 5 mildly  (MKN 3, NGC 3281, NGC 4945, Circinus galaxy and  
IGR J16351-5806) and one heavily (NGC 1068) Compton thick AGN; we therefore estimate the fraction of Compton thick objects to be 
only 7$\%$. 
Although the fraction of absorbed sources is lower than obtained in various Swift/BAT and INTEGRAL/IBIS 
surveys, the percentage of Compton thick AGN is  fully consistent with these previous studies 
(see Table 1 in Ajello 2009).\\

\begin{figure}
\centering
\includegraphics[width=0.8\linewidth]{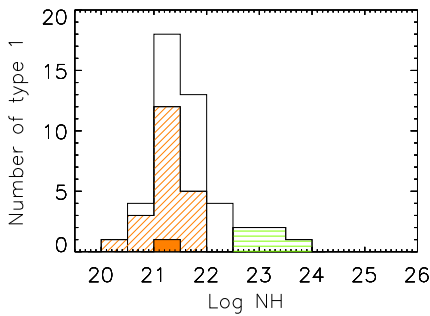}
\caption{Column density distribution in the type 1 objects belonging to the complete sample.
The horizontal dashed bins represents sources requiring complex absorption  for which the higher value 
of N$_{H}$ has been used (see table 1) The diagonally dashed bins represent sources for which upper 
limits on the parameter are available and filled bin is relative again to GRS 1734-292, the only one with a
lower limit estimate on N$_{H}$.}
\label{fig5}
\end{figure}

\begin{figure}
\centering
\includegraphics[width=0.8\linewidth]{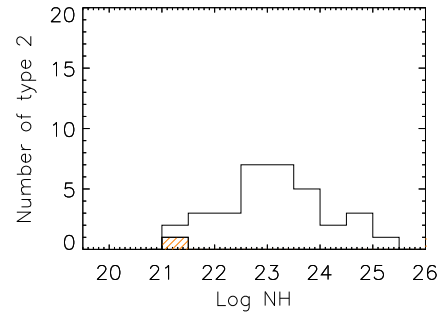}
\caption{Column density distribution in the 33 type 2 AGN of the complete sample.
Dashed bin represents IGR J16024-6107 where no absorption in excess of the Galactic one
has been measured.}
\label{fig6}
\end{figure}

To better investigate the absorption properties of our sample, we have also studied the N$_{H}$ distribution for
type 1 and type 2 objects independently.  We have not considered the column density distribution of blazars since all
but one of these sources have Log(N$_{H}$) values below 21.6 indicative of  small intrinsic absorption. 
The only exception, PKS 1830-211, is a peculiar case of a gravitationally lensed system in which it is not obvious 
where the absorption comes from (De Rosa et al. 2005); besides  the deficit of low energy photons seen in many blazars 
could  equally be ascribed to absorption or to intrinsic spectral curvature (Masetti et al. 2008a) which complicates the issue 
of a column density distribution in these sources.\\
Figure 5 shows the histogram of the column density values measured  in type 1 AGN; upper (diagonal dashed bins) and lower limits
(filled bin) as well as complex absorption values (horizontal dashed bins) were treated separately from measured values.
Almost 20\% of the Seyfert  1 galaxies in our complete sample, have  N$_{H}$ $\ge$ 10$^{22}$ cm$^{-2}$. Of these absorbed
objects, half require complex absorption (i.e. one or more layers of absorbing material partially or fully covering the source); 
these objects lie in the tail of the column density distribution towards high values
(see also Molina et al. 2009). Note however that in these AGN, the value of
the reported column density is that of the layer with the highest
N$_{H}$. \\
Figure 6 shows instead the column density distribution of type 2 objects.
A peak at around Log(N$_{H}$)=23 characterises  this distribution  with all (85\%)  but 5 objects (NGC1275, IGR J07565-4139, IGR J14415-5542, 
IGR J16024-6107 and IGR J17513-2011)
having Log(N$_{H}$)$>$22; the fraction of Seyfert 2  with  Log (N$_{H}$)$>$23 is  55\%  while that of Compton thick objects  is only  18\%.

\section{Comparison with optically selected samples}
By extracting a subsample of objects selected in [OIII] 5007 $\AA$, assumed to trace the intrinsic AGN flux, Risaliti et al. 
(1999) were able to determine the first unbiased N$_{H}$ distribution of  Seyfert 2s  and their paper is 
still used nowadays as a reference work for AGN absorption issues. 
In this sample of optically obscured Seyfert nuclei, the fraction of objects with Log(N$_{H}$)$>$23 is 75\% 
while Compton thick sources are 50\%; i.e. they are as numerous as  Compton thin AGN. We have taken this same sample 
and updated the N$_H$ measurements finding more recent X-ray measurements for many objects and the first absorption 
estimates for five sources.
Our analysis yields a Compton thick fraction of 36\% (15 of 41), slightly lower than that found initially, 
but still considerably higher than found in typical gamma-ray surveys.

It is possible that in our survey we have not recognised some Compton thick AGN because of the low statistial quality
of the X-ray observations used to estimate N$_{H}$.
To see if this has happened we can use the diagnostic diagram provided by Malizia et al. (2007).
This diagram uses  the N$_{H}$ versus  softness ratio (F$_{2-10~keV}$/F$_{20-100~keV}$) to look for AGN candidates 
and its validity has recently been confirmed by Ueda et al. (2007) and  Malizia et al. (2009): 
Misclassified Compton thick objects populate the part of the diagram with low absorption and low softness ratios. 
Figure  7 shows this diagnostic tool applied to our complete sample;  it is evident that only two sources are found in the region of 
Compton thick candidates. Both are type 1 AGN (the blazar 4C 04.42 and the Seyfert 1 NGC 6814) and their location in the diagram 
is probably  accidental; the absorption in the blazar is debatable as it could also be due to intrinsic curvature in the source spectral 
energy distribution (De Rosa et al. 2008b) while the Seyfert  galaxy is known to be variable over time so 
that the low softness ratio is likely due to non-simultaneous X/gamma-ray data (Molina et al. 2006).
We therefore conclude that all Compton thick AGN in our sample have been recognised and properly accounted for.

\begin{figure}
\centering
\includegraphics[width=0.8\linewidth]{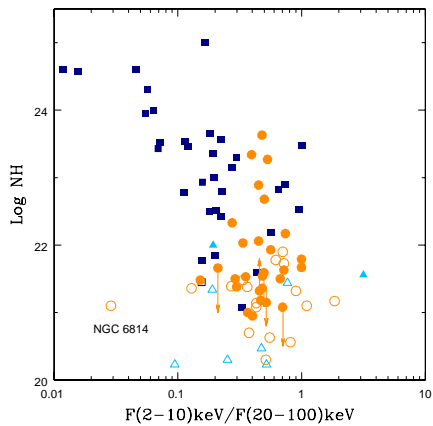}
\caption{Column density plotted against the F$_{2-10~keV}$/F$_{20-100~keV}$ flux ratio of our complete sample of AGN.
Circles are type 1 objects, squares are type 2 and triangles are blazars. Open symbols are objects where no intrinsic 
absorption have been measured.}
\label{fig7}
\end{figure}

We have also verified that our sample, when viewed in OIII, is not significantly different to that of Risaliti et al. 
To this end, we have collected from the literature the [OIII] 5007 $\AA$ fluxes for all our type 2 objects.
As noted by Maiolino and Rieke (1995) the host galaxy gaseous disk might obscure part of the narrow line region where  
the [OIII] 5007 $\AA$ emission originates. To correct for this effect we have used the prescription of Bassani et al. (1999) 
using the observed [OIII] 5007 $\AA$ fluxes and Balmer decrement  H$_{\alpha}$/H$_{\beta}$ and  when the latter
was not available we based our correction on the H$_{\beta}$/H$_{\gamma}$ ratio (see Gu et al. 2006).
For each Seyfert 2, Table 2 lists  the Balmer decrement,  the  corrected [O III] 5007 $\AA$ flux and the related reference; only in one case 
(IGR J20186+4043) are these data missing.\\
In figure 8 the distribution of [O III] 5007$\AA$ fluxes for  our sample (dashed bins) is compared with that of  Risaliti et al. (1999): 
no difference is evident from the figure indicating that we are likely  sampling the same population.\\
 
\begin{figure}
\centering
\includegraphics[width=0.8\linewidth]{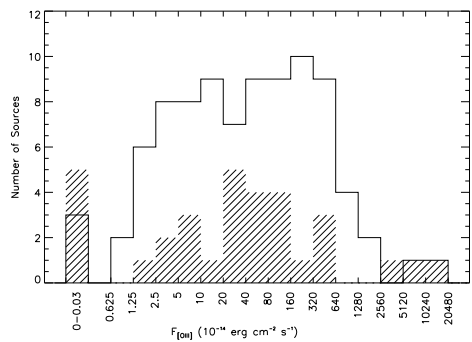}
\caption{[O III] flux distribution of the Risaliti et al. sample compared to the one in INTEGRAL complete
sample (dashed bins)}
\label{fig8}
\end{figure}

The most reasonable explanation for the difference in the fraction of Compton thick objects found in gamma and 
optically selected samples is due to bias introduced by obscuration which reduces the source luminosity by an 
amount depending on the column density. It is therefore more likely that, at a given distance, the most heavily 
absorbed AGN will have a flux below our sensitivity limit than unabsorbed ones and therefore will be lost from our sample.

A method of investigating the number of these 'missing' Compton thick sources is to calculate the reduction in the 
20-40 keV flux as a function of N$_H$ using a simple absorbed power-law model in XSPEC. 
The average flux reduction is negligible below N$_H$=24 and becomes progressively more important thereafter 
(8\%, 25\% and 64\% reduction in the ranges 24-24.5, 24.5-25, and 25-25.5 respectively). 
Despite the simplicity of the fit adopted, the numbers do not change significantly for more complex models.
Starting from the source numbers shown in Figure 6, we can calculate that this reduction in flux would lead 
to the 'loss' of around 15 sources in the Compton thick regime assuming a Euclidian LogN/LogS. 
This suggests that the true fraction of Compton thick sources among Seyfert 2 is around 40\% in reasonable agreement with that 
found for the Risaliti et al. (1999) sample.

Another manner in which to examine the effect of absorption on source numbers is to calculate the fraction of absorbed 
(N$_{H}$$\geq$ 10$^{22}$ cm$^{-2}$) objects compared to the total number of AGN (i.e. the number of objects with 
N$_{H}$$\leq$ 10$^{25}$ cm$^{-2}$) as a function of redshift.
We divided our sample into 5 bins of redshift (up to z=0.335) chosen in order to have a reasonable number of sources in each bin. 
The result is shown in figure 9 where there is a clear trend  of decreasing fraction of absorbed objects 
as the redshift increases.
 
We interpret this evidence as an indication that in the low redshift bin we are seeing almost the entire AGN population, 
from unabsorbed to at least mildly Compton thick; while in the total sample we lose the heavily absorbed 'counterparts' 
of distant and therefore dim sources with little or no absorption.

It is then incorrect to look at the overall sample in order to estimate the role of absorption and one manner in which 
we can come closer to the true picture is by just adopting the first redshift bin for our estimates. 
Despite the lower statistics, we are now in the position to compare our result with that of Risaliti 
et al. in a more correct way.
To do this, we use only the Seyfert 2's in our first redshift bin and then compare their column density distribution 
with that of all type 2 AGN in the Risaliti et al. sample having z$\le$ 0.015. 
Up to this redshift,  there are 17 objects in our sample compared to 39 in that of Risaliti et al. 
Figure 10 shows the results of this comparison: the similarity between the two distributions is 
striking  with the fraction of objects having N$_{H}$$\geq$ 10$^{23}$ cm$^{-2}$
being similar in the two samples ($\sim$ 75\%). The fraction of Compton thick objects is also remarkably close 
(35\% compared to 36\%).

In conclusion every method we use leads to an estimate of around 36\%-40\% for  the true fraction of Compton 
thick AGN among Seyfert 2.
Going from just the Seyfert 2 to the entire AGN population we note that 
the first bin, ranging up to z = 0.015, contains 25 AGN, of which 20 (80\%)  are absorbed and of these, 6 (24\%)
are Compton thick. 
It is still possible that the measured fraction of Compton 
thick objects is a lower limit, since some of the most heavily absorbed  sources may  not have  sufficient luminosities 
to be detected even at the lowest redshifts.\\

We have also analysed the sources in the first redshift bin to look for a trend of decreasing  
fraction of absorbed AGN with increasing  source gamma-ray luminosities. 
This effect, which is well documented in the X-ray band (La Franca et al. 2005), has also been observed in gamma-rays 
(Bassani et al. 2006, Sazonov et al. 2007 and references therein) although the redshift effect discussed 
here may have contaminated the result. 
Dividing the 25 sources with z$\le$0.015 into two luminosity bins,  we find comparable fractions of absorbed sources.
This means that either our statistics  are  too low
for a proper estimate or the effect is not real but only induced by the selection due to z.

\begin{small}
\begin{figure}
\centering
\includegraphics[width=0.8\linewidth]{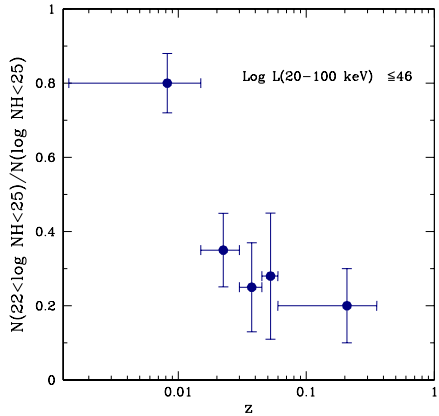}
\caption{Fraction of absorbed  objects compared to the total number of AGN  as a function of redshift.}
\label{zeta}
\end{figure}
\end{small}

\begin{small}
\begin{figure}
\centering
\includegraphics[width=0.8\linewidth]{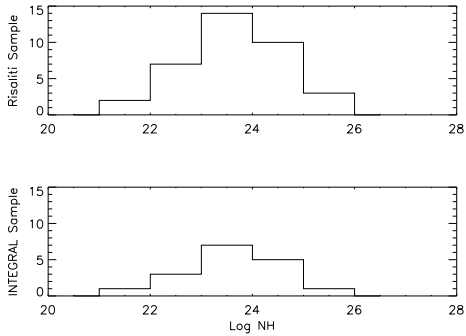}
\caption{Comparison of the distribution of column densities in the type 2 objects between Risaliti et al. sample (up) and 
INTEGRAL sample (bottom) with z$\leq$0.015. }
\label{fig10}
\end{figure}
\end{small}

\begin{table}
\begin{center}
\centerline{Table 2: OIII fluxes of the Type 2 objects}
\label{table}
 \begin{tabular}{l l l  r}
  \hline\hline
    Name               &   H$_{\alpha}$/H$_{\beta}$ &  F$_{O III}^{\dagger}$  & Ref  \\
\hline
NGC 788                       &   1.82$^{\star}$           & 3.65                 & 1 \\
NGC 1068                     &   7.00                           & 15860              & 2 \\
NGC 1142                     &   3.37                           & 1.59                  & 3  \\
NGC 1275                     &   5.00                           & 31.1                  & 2 \\
Mkn 3                             &   6.67                           & 4600                 & 2 \\
IGR J07565-4139        &   $>$11.2                    & $<$20                & 4 \\
MCG-05-23-16             &   8.00                          & 41                       & 2 \\
IGR J09523-6231        &    (a)                            &  $\geq$0.2        & 5 \\
SWIFT J1009.3-4250  &   4.74                           &  20.72              & 6 \\
NGC 3281                     &   6.13                          & 45                     & 2 \\
IGR J10404-4625        &   $<$17.14                 & $<$454           & 5 \\
IGR J12026-5349        &   9.2                             & 245                  & 4\\
NGC 4388                      &   5.50                         & 374                  & 2 \\
NGC 4507                     &   4.50                          & 158                  & 2  \\
LEDA 170194              &   6.65                           & 62.22              & 7 \\
NGC 4945                       &     -                             & $>$40             & 8 \\
IGR J13091+1137        &    (b)                           & $>$0.17         & 9 \\
Cen A                             & 5.50                            & 7                    & 2 \\
NGC 5252                     &   3.72                         & 39                  & 2 \\
Circinus Galaxy            &   19.1                        & 6970             & 2 \\
NGC 5506                    &   7.20                         & 600               & 2 \\
IGR J14515-5542        &   $>$15                    &  20.4            & 4 \\
IC 4518A                        &   5.5                         &  105             & 5 \\
IGR J16024-6107        &   5.33                        & 5.47            & 5 \\
IGR J16351-5806        &   2.24$^{\star}$        &  33             & 1 \\
NGC 6300                      &  2.58$^{\star}$       & 7.44           & 1 \\
IGR J17513-2011        &  (a)                            &  $>$0.15    & 4 \\
ESO 135-G35               &   6.31                        & 112               &   2 \\
Cyg A                             &   5.40                        & 80              & 2 \\
NGC 7172                     &   3.00                     & 4.0                  & 2 \\
IGR J23308+7120        &    (c)                     & $>$0.27            & 5 \\
IGR J23524+5842        &    (d)                     & $>$0.06            & 5 \\
\hline
\end{tabular}
\end{center}
Notes: ($\dagger$) = corrected line flux in units of 10$^{-14}$ erg cm$^{-2}$ s$^{-1}$, $^{\star}$  H$_{\beta}$/H$_{\gamma}$;
(a): H$_{\alpha}$ heavily blended with [N II] lines; (b): H$_{\beta}$ not detected; (c): H$_{\beta}$ in absorption; 
(d): H$_{\alpha}$ not detected .
Ref.: (1) Gu et al. 2006; (2) Bassani et al. 1999; (3) Moustakas and Kennicutt 2006; (4) Masetti et al. 2006a; 
(5) Masetti et al. 2008b;
(6) Landi et al. 2007a; (7) Masetti et al. 2006b; (8) Risaliti et al. 1999; (9) Masetti et al. 2006c.
\end{table}

\section{Summary and Conclusions}
In this work a complete sample of soft gamma-ray (20-40 keV) selected AGN  has been extracted by means of 
the V$_{e}$/V$_{a}$ test and this has been used to study the absorption in the local Universe.
As expected the N$_{H}$ distribution turns out to be quite different in the two classes of type 1 and type 2 objects
with the first peaking at Log(N$_{H}$)=21-22 and the second at Log(N$_{H}$)=23.
If we consider N$_{H}$ $\geq$ 10$^{22}$ cm$^ {-2}$ as the dividing line between absorbed 
and unabsorbed sources, we find that absorption is present in 43\% of the sample with only 6 objects being Compton thick 
AGN i.e. a fraction of 7\%.
Taking into account this and previous high energy surveys, the Compton thick sky  currently sampled above 10 keV, contains a 
low number of heavily obscured (N$_{H}$ $\geq$ 10$^{24}$ cm$^ {-2}$) objects (see also Ajello 2009), significantly less than 
those found in the optically selected sample of Risaliti et al. (1999). \\
We have shown that this could be due to the observed flux of many Compton thick AGN being reduced by obscuration to a 
level below our detection limit.
Furthermore we find evidence of a selection effect due to redshift at high energies which may well reconcile the results obtained 
from gamma-ray and optically selected samples.
Dividing our sample into 5 bins of $z$, up to 0.335,  we find a clear trend for a decreasing fraction of absorbed 
objects as the redshift increases, in particular  in the first bin (up to $z\leq$0.015), containing 25 AGNs, 80\%  of the sources 
are absorbed.

Furthermore, a comparison of the Seyfert 2 objects in our sample and in the Risaliti et al. sample up to $z$ $\leq$0.015,
provides a similar column density distribution: the fraction of AGN with N$_{H}$ $\geq$ 10$^{23}$ cm$^ {-2}$ are
equal ($\sim$ 75\%) in the two samples, as is the fraction of Compton thick objects ($\sim$ 35\%).

It is now possible to correct our results for any bias due to $z$, i.e. by analysing only 
sources located within $\sim$ 60 Mpc. Within this set of sources the fraction of Compton thick objects is 24\%.
Future IBIS and BAT surveys will provide a larger database of nearby AGN allowing a confirmation of this result and more in depth analysis
of absorption in the local Universe.

\section*{Acknowledgements}
The authors would like to thank Dr. Nicola Masetti for his help in collecting the [O III] 5007 $\AA$ fluxes and the anonymous referee for 
suggestions  on how to improve the manuscript.\\
We acknowledge ASI  financial and programmatic support via contracts I/008/07/0.

\end{document}